\documentclass{article}
\pdfoutput=1

\usepackage{jheppub}
\usepackage{amsmath}
\usepackage{amssymb}
\usepackage{booktabs}
\usepackage{braket}
\usepackage{colortbl} 
\usepackage{comment}
\usepackage{dcolumn}
\usepackage{graphicx}
\usepackage[utf8]{inputenc}
\usepackage{multirow}
\usepackage{siunitx}
\usepackage[normalem]{ulem}
\usepackage{xcolor}
\usepackage{xspace}
\usepackage[nameinlink]{cleveref}
\usepackage{physics}
\usepackage{cancel}
\usepackage{slashed}

\usepackage{tikz}
\usetikzlibrary{positioning,decorations.pathreplacing}
\usepackage[compat=1.1.0]{tikz-feynman}
\tikzfeynmanset{warn luatex=false}

\allowdisplaybreaks

\newcommand{\eg}{\textit{e.g.}\xspace}
\newcommand{\ie}{\textit{i.e.}\xspace}
\newcommand{\GeV}{\textrm{GeV}\xspace}
\newcommand{\MeV}{\textrm{MeV}\xspace}
\newcommand{\GFermi}{\ensuremath{G_\textrm{F}}\xspace}

\renewcommand{\Im}{\operatorname{Im}}

\renewcommand{\TH}{T_\text{H}}
\newcommand{\Thom}{T_\text{hom}}
\newcommand{\Tinhom}{T_\text{inhom}}
\newcommand{\TFSR}{T_\text{FSR}}
\newcommand{\npq}{\ensuremath{(n_+ q)}}
\newcommand{\nmq}{\ensuremath{(n_- q)}}

\begin{document}

\title{On soft contributions to the $B^-\to \gamma^*$ form factors}
\author[a]{Aoife Bharucha,}
\emailAdd{aoife.bharucha@cpt.univ-mrs.fr}
\author[b]{Danny van Dyk,}
\emailAdd{danny.van.dyk@gmail.com}
\author[a]{Eduardo Velásquez}
\emailAdd{edvelasquezal7@gmail.com}

\affiliation[a]{Aix Marseille Univ, Université de Toulon, CNRS, CPT, Marseille, France}
\affiliation[b]{Institute for Particle Physics Phenomenology and Department of Physics, Durham University, Durham DH1 3LE, UK}

\abstract{
The photoleptonic decay $B^-\to \gamma \ell^-\bar{\nu}$ is the simplest low-energy process that probes the substructure of the $B$ meson,
making it an excellent candidate to determine the parameters of $B$-meson light-cone distribution amplitudes from experimental data.
More recently, the decay $B^-\to \gamma^*(\to \ell'^-\ell'^+)\ell^-\bar\nu$ has received attention as an alternative probe of these parameters.
Both decays are described through a common set of hadronic form factors, which if computed within the framework
of QCD factorization give rise to the sensitivity to the $B$-meson light-cone distribution amplitudes.
Nevertheless, in this case the form factors still receive so-called soft contributions that can only be estimated but not rigorously computed.
In this work, we provide results for the QCD factorization expressions for the form factors, including terms at next-to-leading
power in the $b$ quark mass and in the photon energy. We further use these results to estimate the soft contributions
within a light-cone sum rule setup.
Finally, using numerical results obtained from a benchmark model of the light-cone distribution amplitudes,
we show that the soft contributions are under significantly better theoretical control at a mildly spacelike photon virtuality.
}

\begin{flushright}
    IPPP/26/09
\end{flushright}
\vspace*{-3\baselineskip}

\maketitle

\section{Introduction}
\label{sec:intro}

The phenomenology of $B$-meson decays is an important source of information for third-generation Standard Model (SM) couplings, such as the
Cabibbo-Kobayashi-Maskawa matrix elements $V_{ub}$ and $V_{cb}$~\cite{BaBar:2014omp}. Moreover, it provides strong constraints on Physics beyond the SM,
\eg, through measurements of rare $B\to K^*\mu^+\mu^-$ decays; see \eg~\cite{Albrecht:2021tul}. Common to both cases of $B$-meson phenomenology is the need to control
decay-specific hadronic matrix elements. The latter are commonly expressed in terms of form factors, \ie~scalar valued
functions of the relevant non-trivial kinematic variables.
Beside lattice QCD simulations~\cite{FlavourLatticeAveragingGroupFLAG:2024oxs}, an important source of information on these form factors is the framework of
QCD factorization (QCDF), also called collinear factorization~\cite{Beneke:2000wa,Beneke:2002ph,Beneke:2001at,Beneke:2004dp}.
Another source are light-cone sum rules with an on-shell $B$ meson~\cite{Khodjamirian:2005ea,Khodjamirian:2006st,Faller:2008tr,Gubernari:2018wyi}.
Beyond the form factors, QCDF calculation of nonleptonic $B$ decays provide important phenomenological inputs~\cite{Beneke:1999br,Beneke:2000ry,Beneke:2001ev}.
\\

A dominant source of uncertainty for calculations within the latter three frameworks
emerges from our incomplete knowledge of the $B$ meson light-cone distribution amplitudes (LCDAs).
A particularly important role in such calculations is played by the inverse moment $\lambda_B^{-1}$ of the twist-2 LCDA  $\phi_+(\omega,\mu)$, defined by (see \eg~ref.~\cite{Braun:2003wx})
\begin{equation}
	\label{eq:intro:lambdaB}
	\lambda_B^{-1}(\mu)=\int_{0}^{\infty} \frac{d\omega}{\omega}\phi_+(\omega,\mu)\,.
\end{equation} 
Here, $\omega$ corresponds to a light-cone projection of the spectator quark momentum, and $\mu$ represents the renormalisation scale of the LCDA
and its inverse moment. Further inverse-logarithmic moments arise in QCDF calculations at next-to-leading order in $\alpha_s$.
The light-cone distribution amplitude $\phi_+$ (and its cousins discussed below) are defined within the framework
of the heavy quark effective theory (HQET).
Neither the inverse moment, its logarithmic cousins, nor the functional form of the LCDA $\phi_+$ are currently known from first principles.
The prevalent approach to extract information on these quantities, and therefore the substructure of the $B$ meson, involves studying the simplest possible
process that involves a $B$-meson process and that is sensitive to the $B$-meson LCDAs.
The photoleptonic decay $B^-\to \gamma \ell^-\bar{\nu}$ is an excellent candidate for this purpose.
For an energetic photon, its amplitude can be accessed within the framework of QCDF
~\cite{Beneke:2000wa,Grozin:1996pq,Bosch:2003fc,Descotes-Genon:2002crx}.
The sensitivity to the LCDA parameters in analyses involving the photoleptonic decay has been recently studied in detail~\cite{Lughausen:2024pmk}.

The physical photoleptonic amplitude arises in the $q^2 \to 0$ limit of the $B^-\to \gamma^*\ell^-\bar\nu$ amplitude,
where $q$ denotes the photon momentum. The amplitude involving an off-shell photon is relevant to the description of
the decay $B^-\to \ell'^+\ell'^- \ell^-\bar\nu$, which was first studied in the context of QCD factorization in refs.~\cite{Bharucha:2021zay,Beneke:2021rjf,Wang:2021yrr}.  The process with an off-shell photon provides an alternative manner to access the LCDAs, being more accessible both at LHCb and Belle II, and results in a larger set of observables~\cite{Wang:2021yrr}.
However, for timelike photon virtuality, the treatment of hadronic resonances in the photon spectrum are a challenge~\cite{Beneke:2021rjf}.
In this work, we investigate the general case of an off-shell photon; the on-shell limit is discussed where relevant.

The description of both of these processes involves the hadronic two-point correlation function
\begin{equation}
    \label{eq:intro:def-TH}
    \TH^{\mu\nu}(k,q)
        = \int dx \ e^{i q\cdot x} \matrixel{0}{T\{J_\text{em}^\mu(x) J_\mathrm{H}^\nu(0)\}}{B^-(q + k)}\,,
\end{equation}
where we introduce $k^\mu$ as the lepton-neutrino pair momentum.
Moreover,
\begin{equation}
    \label{eq:intro:Jem}
    J_{\mathrm{em}}^\mu(x)=\bar{q}(x) \mathcal{Q} \gamma^\mu q(x)+\sum_{\ell} Q_{\ell} \bar{\ell}(x) \gamma^\mu \ell(x)
\end{equation}
represents the electromagnetic current. We abbreviate $q(x)=(u(x), d(x), s(x), c(x), b(x))^{\mathrm{T}}$ and introduce the
lepton charge $Q_\ell = -1$ and the quark charge matrix $\mathcal{Q}=\operatorname{diag}[2 / 3,-1 / 3,-1 / 3,2 / 3,-1 / 3]$.
Furthermore,
\begin{equation}
    \label{eq:intro:JH}
    J_\mathrm{H}^\nu=\bar{u}(x)\gamma^\nu(1-\gamma_5)b(x)
\end{equation}
represent the flavour-changing $b\to u$ current.
As the dominant contribution to \cref{eq:intro:def-TH} arises at lightlike distances $x^2 \simeq 1/(m_b \Lambda_\text{had})$,
this introduces a dependence on the $B$-meson LCDAs. Corrections to the dominant light-cone contribution $x^2 = 0$ arise
as operators of higher-twist in a light-cone operator product expansion.
However, despite this light-cone dominance, soft configurations also contribute to the time-ordered product via field configurations with $x^2 \simeq 1/(\Lambda_\text{had}^2)$.
The soft contributions are not accessible within the framework of QCDF. An estimate using a light-cone sum rule setup was first discussed
for the case of an on-shell photon in ref.~\cite{Braun:2012kp}; see also refs.~\cite{Ball:2003fq, Wang:2018wfj} for a sum rule setup using the photon LCDAs.

The current state of the art calculation of the $B^-\to \gamma$ form factors in QCD factorization include the leading twist contribution at NNLL accuracy~\cite{Beneke:2011nf}, next-to-leading power corrections and higher-twist corrections up to twist 6~\cite{Wang:2016qii,Beneke:2018wjp}.
The calculation of the soft contributions in terms of the twist-2 LCDAs amplitude at leading order in $\alpha_s$ was proposed
in ref.~\cite{Braun:2012kp}; further refinements including the $\mathcal{O}(\alpha_s)$ and higher-twist corrections followed in refs.~\cite{Wang:2016qii,Wang:2018wfj,Beneke:2018wjp}.
The extension to the off-shell photon case was first calculated in QCDF in refs.~\cite{Bharucha:2021zay,Beneke:2021rjf,Wang:2021yrr}.
Afterwards, a new form factor basis was proposed that avoids kinematic singularities, which is a prerequisite for the application of dispersion relations~\cite{Kurten:2022zuy}.

In this paper, we calculate the off-shell form factors in this complete and singularity-free form factor basis within the framework of QCDF, including higher-power corrections and higher-twist contributions up to twist 4. 
We further revisit the derivation of the original light-cone sum rule setup for an estimate of the soft contributions and apply it to the off-shell case.
The manuscript is structured as follows: in \cref{sec:framework:twist-2} we introduce the QCD factorization framework. We then derive results for the higher-twist contributions to the form factors in a dispersive form in \cref{sec:framework:higher-twist}.
In \cref{sec:framework:soft-cont}, we estimate the soft contributions from our previous results for the QCDF integration kernels up to the
twist-four level.
We then present our numerical results for the ratio of the soft to the QCDF contributions in \cref{sec:numerics} and our conclusions in \cref{sec:conclusions}.

\section{Factorization framework and beyond}
\label{sec:framework}

To leading order in QED, the photoleptonic amplitude reads
\begin{equation}
    \label{eq:framework:def-amp}
    \mathcal{M}(B^-(p)\to\ell^-(p_\ell)\bar{\nu}_\ell(p_\nu)\gamma^*(q, \varepsilon)) =
        \frac{4 \GFermi V_{ub}}{\sqrt{2}}
        \bra{\ell^-\bar{\nu}_\ell\gamma^*}
            J^\nu_\mathrm{L}(0)J_{\mathrm{H}\nu}(0)
        \ket{B^-}\,.
\end{equation}
The prefactor contains the Fermi constant $\GFermi$ and the Cabibbo-Kobayashi-Maskawa matrix element $V_{ub}$.
We abbreviate the charged currents as $J^\nu_\mathrm{L}(x)=\bar{\ell}(x)\gamma^\nu(1-\gamma_5)\nu_\ell(x)$
and $J_{\mathrm{H}\nu}$, the leptonic and hadronic weak currents, respectively; see \cref{eq:intro:JH} for the definition of the latter.
It is convenient to split the matrix element into a hadronic tensor $\TH$ and a final-state radiation (FSR)
tensor $\TFSR$~\cite{Kurten:2022zuy}:
\begin{equation}
    \label{eq:framework:def-TH+TFSR}
    \bra{\ell^-\bar{\nu}_\ell\gamma^*}
        J^\nu_\mathrm{L}(0)J_{\mathrm{H}\nu}(0)
    \ket{B^-} =
    e Q_{B^-} \varepsilon^*_\mu
    \left[\TH^{\mu\nu}(k, q)
        + \TFSR^{\mu\nu}(p_\ell, p_\nu)
    \right] L_\nu\,.
\end{equation}
Here $\TH^{\mu\nu}(k, q)$ is the hadronic tensor defined in~\cref{eq:intro:def-TH}, $e$ is the positron charge, $Q_{B^-} = -1$ is the $B$-meson's relative charge in units of $e$, $\varepsilon^\mu$ is the polarisation vector of the photon, and we abbreviate $L^\nu = \bra{\ell^-\bar\nu} J_{L}^\nu \ket{0}$.
The second term, $\TFSR$, can be computed perturbatively to order $\alpha_e^1$ by taking into account the FSR
off the charged-lepton; we refer to \cref{app:details:FSR} for its definition.
The splitting of the matrix element into the two terms $\TH$ and $\TFSR$ is somewhat arbitrary (albeit convenient) and certainly not gauge invariant;
hence, the two tensors do not independently fulfil the Ward identity. Following ref.~\cite{Kurten:2022zuy}, we further split the hadronic tensor into
a homogeneous and non-homogeneous part, $\TH = \Thom + \Tinhom$; see \cref{app:details:singularity-free-hadronic-tensor} for their definitions.
The former admits a Lorentz decomposition in terms of four scalar-valued functions:
the $B^-\to \gamma^*$ form factors $F_1$ through $F_4$~\cite{Kurten:2022zuy}\footnote{%
    Note that, despite identical naming, this basis of form factors should not be confused for the form factors
    arising in ref.~\cite{Wang:2021yrr}.
}.
The latter is fixed by demanding that its sum with $\TFSR$ restores gauge invariance and that it does
not introduce spurious kinematic singularities in $q^2$ in the form factors $F_1$ to $F_4$.
The absence of such kinematic singularities is a formal prerequisite for expressing these form factors
through hadronic dispersion relations~\cite{Kurten:2022zuy}.
The decomposition of $\Thom$ in terms of this complete and singularity-free basis of form factors reads
\begin{equation}
    \label{eq:framework:def-FFs}
    \begin{aligned}
        \Thom^{\mu\nu}(k, q) 
            & = \frac{1}{M_B}\left[(k \cdot q) g^{\mu \nu}-k^\mu q^\nu\right] F_1\left(k^2, q^2\right) \\
            & \phantom{=}\,\, + \frac{1}{M_B}\left[\frac{q^2}{k^2} k^\mu k^\nu-\frac{k \cdot q}{k^2} q^\mu k^\nu+q^\mu q^\nu-q^2 g^{\mu \nu}\right] F_2\left(k^2, q^2\right) \\
            & \phantom{=}\,\, + \frac{1}{M_B}\left[\frac{k \cdot q}{k^2} q^\mu k^\nu-\frac{q^2}{k^2} k^\mu k^\nu\right] F_3\left(k^2, q^2\right) \\
            & \phantom{=}\,\, + \frac{i}{M_B} \epsilon^{\mu \nu \rho \sigma} k_\rho q_\sigma F_4\left(k^2, q^2\right)\,,
    \end{aligned}
\end{equation}
where $M_B$ is the $B$-meson mass.
The form factors can be extracted from $\Thom^{\mu\nu}(k, q)$ in unambiguous way by contracting it with the projectors $P_1$ through
$P_4$ as detailed in \cref{app:details:projectors}.
This choice of form factor basis can be related to the common decomposition~\cite{Beneke:2021rjf}, which involves only three
form factors: one vector form factor $F_V$, and two axial form factors $F_{A_\perp}$ and $F_{A_\parallel}$.
The relationships read
\begin{equation}
    \label{eq:framework:relations-to-Beneke}
    \begin{aligned}
        F_V & = F_4\,,\\
        F_{A_\perp} & = \frac{(k\cdot q) F_1(k^2, q^2) - q^2 F_2(k^2, q^2)}{(k + q)\cdot q}\,,\\
        F_{A_\parallel} & = q^2\left[\frac{F_1(k^2, q^2) + F_2(k^2, q^2)}{(k + q)\cdot q}+\frac{F_2(k^2, q^2)-F_3(k^2, q^2)}{k^2}\right]\,.
    \end{aligned}
\end{equation}
In the limit $q^2\to 0$ the form factors in \cref{eq:framework:relations-to-Beneke}
become $F_{A_\perp}(k^2) = F_1(k^2, 0)$ and $F_{A_\parallel}(k^2) = 0$.
A fourth pseudoscalar form factor is not defined in the common decomposition, since it is not of phenomenological relevance
for decays involving light-leptons: it does not contribute in the limit $m_\ell \to 0$~\cite{Kurten:2022zuy}.

\subsection{Leading factorizable contributions}
\label{sec:framework:twist-2}

For an on-shell photon with large energy in the $B$-meson rest frame,
the time-ordered product in \cref{eq:intro:def-TH} is dominated by field configurations at lightlike distances, 
\ie~$x^2 = 0$~\cite{Descotes-Genon:2002crx,Beneke:2011nf}.
This light-cone dominance is most readily discussed using light-cone coordinates. To this end, one commonly
introduces two lightlike vectors $n_+$ and $n_-$, with $n_+ \cdot n_- = 2$, and write
\begin{equation}
\label{eq:framework:twist-2:kinematics}
\begin{aligned}
    p_b^\mu & = \frac{m_b}{2} \left(n_+ + n_-\right)^\mu = m_b v^\mu\,, &
    p_s^\mu & = \frac{\omega}{2} n_+^\mu\,,\\
    q^\mu   & = \frac{\npq}{2} n_-^\mu + \frac{\nmq}{2} n_+^\mu + q_\perp^\mu\,, &
    k^\mu   & = \frac{m_b - \npq}{2} n_-^\mu + \frac{m_b - \nmq}{2} n_+^\mu - q_\perp^\mu\,.
\end{aligned}    
\end{equation}
Here, $p_b$ and $p_s$ are the momenta of the $b$ quark and the spectator anti-quark, respectively.
Light-cone dominance $x^2 \lesssim 1/(m_b \Lambda_\text{had})$ is achieved if $\npq = \order{m_b}$
and $\nmq = \order{\Lambda_\text{had}}$.

As a consequence of the light-cone dominance, a light-cone expansion of the relevant propagators becomes possible,
leading to a systematic expansion of the form factors in terms of $\Lambda_\text{had}/m_b$,
$\Lambda_\text{had}/(n_+ q)$, and $\alpha_s$. The leading power result for each form factor
factorizes into hard-collinear scattering kernels $T_i$ and the leading two-body light-cone distribution
amplitude of the $B$ meson $\phi_+(\omega;\mu)$:
\begin{equation}
    \label{eq:framework:twist-2:def-hc-kernels}
    F_i(q^2, k^2)
    = \frac{f_B}{n_+q} \int_0^\infty d\omega \ T_{i}^{\text{tw2}}(\omega, n_+q, n_-q;\mu)\, \phi_+(\omega;\mu) + \order{\frac{\Lambda_\text{had}}{\lbrace n_+q, m_b\rbrace}}^2\,.
\end{equation}
Here $\omega$ is the usual leading light-cone component of the spectator quark momentum.
The above holds for two-body Fock states within the $B$ meson; three-body contributions
have similar factorization formulas as discussed in ref.~\cite{Beneke:2018wjp}.
The factorization formula \cref{eq:framework:twist-2:def-hc-kernels} receives further contributions due to off-light-cone field configurations. Contributions arising from $x^2 \simeq 1/(\Lambda_\text{had} m_b)$ are captured by a systematic expansion in $x^2$,
which involve light-cone distribution amplitudes of higher twist~\cite{Beneke:2018wjp}.

We expand hard scattering kernels for the form factors in \cref{eq:framework:twist-2:def-hc-kernels} to leading power in $\Lambda_\text{had}/m_b$.
Despite the fact that $\omega$ is integrated on the interval $[0, \infty)$,
we assume a power counting of $\omega \sim \Lambda_\text{had}$.
This is motivated by a rapid decline of the leading-twist LCDA $\phi_+(\omega)$
that always multiplies the leading-power two-particle hard scattering kernels. 
In our notation, this yields
\begin{equation}
\label{eq:framework:twist-2:hc-kernels}
\begin{aligned}
    T_{1}^{\text{tw2}} & = C_V^{(A 0)} K^{-1} J\frac{M_B Q_u}{\omega-n_-q}\,,                        &
    T_{2}^{\text{tw2}} & = 2C_V^{(A 0)} K^{-1} J\frac{M_B(M_B - n_+q) Q_u}{(n_+q)(\omega-n_-q)}\,,   \\
    T_{3}^{\text{tw2}} & = 0 + \order{\alpha_s \frac{\Lambda}{m_b}},                                 &
    T_{4}^{\text{tw2}} & = C_V^{(A 0)} K^{-1} J\frac{M_B Q_u}{\omega-n_-q}\,,
\end{aligned}
\end{equation}
where $C_V^{(A 0)}\left(n_+q;\mu\right)=U_1\left(n_+q;\mu_{h1},\mu\right)C_V^{(A 0)}\left(n_+q;\mu_{h1}\right)$
is the Wilson coefficient that arises when the QCD heavy-to-light current $J_H$, defined in \cref{eq:intro:JH},
is matched to the corresponding soft collinear effective theory (SCET) current at the hard scale $\mu_{h1}$; $K^{-1}(\mu)=U_2(\mu_{h2},\mu)K^{-1}(\mu_{h2})$
is the conversion factor between the scale dependent decay constant and the physical $B$ meson decay constant to
one loop accuracy as $f_B=F_B(\mu)K(\mu)$; $J\left(n_+q, n_-q,\omega;\mu\right)$ is the hard collinear matching function.
Here we introduce $U_1\left(n_+q;\mu_{h1},\mu\right)$, the evolution factor associated with the SCET current
and $U_2(\mu_{h2},\mu)$, arising from the matching of the $B$ meson decay constants in QCD with HQET.
Explicit formulae for $C_V^{(A 0)}\left(n_+q;\mu_{h1}\right)$ and $J\left(n_+q, n_-q,\omega;\mu\right)$ are given
in~\cite{Beneke:2021rjf} while those for $K^{-1}(\mu_{h2})$, $U_1\left(n_+q;\mu_{h1},\mu\right)$ and $U_2(\mu_{h2},\mu)$
are given in~\cite{Beneke:2011nf}.
For brevity, we have dropped the arguments of these functions in \cref{eq:framework:twist-2:hc-kernels}.

Our results provide the hard-scattering kernel for the complete and singularity-free basis of physical form factors for the first time.
We confirm the general structure at leading power and to leading order in $\alpha_s$ discussed
in the literature~\cite{Braun:2012kp,Beneke:2018wjp,Beneke:2021rjf}: the hard-scattering kernels
exhibit a universal $\omega$ dependence.

\subsection{Power corrections and higher-twist contributions}
\label{sec:framework:higher-twist}

In this section we will present our results for the form factors at non-zero $q^2$ and including next-to-leading power corrections.
We express these results as 
\begin{equation}
    \label{eq:framework:higher-twist:Fcontributions}
    F_i(v\cdot q, q^2) = F_i\big|_{\text{LP}} + F_i\big|_{1/(v\cdot q)} + F_i\big|_{1/m_b}\,,
\end{equation}
for $i=1,2,4$, separating out the leading-power (LP), $1/(v\cdot q)$ and $1/m_b$ suppressed contributions,
where $v = (n_+ + n_-)/2$ is the velocity of the $B$ meson: $p^\mu = M_B v^\mu$.
Here, we prefer to use $v\cdot q$ in lieu of $k^2$ due to its manifest role in the power expansion.
Note that we will not extract
higher-twist contributions to the form factor $F_3$,  due to its nature as a power-suppressed form factor.
The LP terms are given in \cref{eq:framework:twist-2:def-hc-kernels,eq:framework:twist-2:hc-kernels}. 
As mentioned previously, corrections to the factorization formula \cref{eq:framework:twist-2:def-hc-kernels} arise
due to higher twist light-cone operators. The twist of an operator increases either when its mass dimension increases,
in which case the hard-scattering kernels $T_i$ in \cref{eq:framework:twist-2:def-hc-kernels} receive additional contributions, or when additional fields arise, in which case generalized kernels need to be defined.

Following the approach in ref.~\cite{Beneke:2021rjf}, we introduce the light-cone expansion of the hadronic tensor $T_H$
up to the twist-four level in heavy quark effective theory (HQET).
This leads to a different choice of kinematic variables, \ie~
\begin{equation}
\begin{aligned}
    p_b^\mu & = m_b v^\mu + \tilde{k}^\mu\,, &
    p_s^\mu & = \omega v^\mu\,.
\end{aligned}    
\end{equation}
where $\tilde{k}^\mu$ is a small ``residual" momentum satisfying $\tilde{k}^2\ll m_b^2$. Expanding the light-quark propagator in a soft background gluon field beyond leading order~\cite{Balitsky:1987bk} yields
\begin{equation}
    \label{eq:framework:higher-twist:TH}
    \begin{aligned}
    T_{\text{H},\text{LP}+1/(v\cdot q)}^{\mu\nu}
        = & -\frac{Q_u f_B M_B}{2\pi^2}\int d^4x\ \frac{e^{i q\cdot x}}{x^4}v_\rho x_\sigma\,
            \left(g^{\mu\nu}g^{\rho\sigma}+i\epsilon^{\mu\nu\rho\sigma}-g^{\mu\rho}g^{\nu\sigma}-g^{\mu\sigma}g^{\nu\rho}\right)\\
        & 
            \phantom{-}\,\,\left[\Phi_+(v\cdot x)+x^2G_+^{\text{WW}}(v\cdot x)-\frac{x^2}{2(v\cdot x)^2}\Phi_-^{\text{t3}}(v\cdot x)\right.\\
        &
            \left.\phantom{-}\,\,\phantom{\left(\right.}-\frac{x^2}{4}\int_0^1 du\ \big[\Psi_4-\tilde{\Psi}_4\big]^{\text{t4}}\left(v\cdot x,u(v\cdot x)\right)\right]\,.
\end{aligned}
\end{equation}
In the above, we use $\Phi_+(v\cdot x)$ to denote the two-particle (2pt) leading-twist (tw2) LCDA.
Further LCDAs arise, including $\Phi_-^{\text{t3}}(v\cdot x)$ the two-particle twist-three (tw3) LCDA,
$G_+^{\text{WW}}(v\cdot x)$ the two-particle twist-four (tw4) LCDA, and
$\big[\Psi_4-\tilde{\Psi}_4\big]^{\text{t4}}(v\cdot x,u(v\cdot x))$ as one of the three-particle (3pt) twist-four LCDAs.
These LCDAs follow from the the definitions in ref.~\cite{Braun:2017liq}, which we repeat for convenience in
\cref{app:details:lcda-definitions}.
In the notation used throughout this work, upper-case letters refer to these LCDAs as functions in position space,
while the same letter in lower case denotes its momentum space Fourier conjugate function.

We set out to derive the higher-twist contribution presented in \cref{eq:framework:higher-twist:TH} as dispersive integrals,
which is necessary to estimate their impact in the soft contributions. Before doing so, we note that the three-particle LCDA $\big[\Psi_4-\tilde{\Psi}_4\big]^{\text{t4}}(z_1,z_2)$ cannot be written
in terms of a generic profile function, contrary to the orthogonal linear combination $\big[\Psi_4+\tilde{\Psi}_4\big](z_1,z_2)$~\cite{Beneke:2018wjp}.
Hence, we replace the former with the latter, using \cref{eq:Psi4}. Further eliminating $G_+^{\text{WW}}(v\cdot x)$ using \cref{eq:GWWpl}, the hadronic tensor takes the form
\begin{align}
    T_{\text{H},\text{LP}+1/(v\cdot q)}^{\mu\nu}
        & = -\frac{Q_u f_B M_B}{2\pi^2}\int d^4x\ \frac{e^{i q\cdot x}}{x^4}v_\rho x_\sigma\,
            \left(g^{\mu\nu}g^{\rho\sigma}+i\epsilon^{\mu\nu\rho\sigma}-g^{\mu\rho}g^{\nu\sigma}-g^{\mu\sigma}g^{\nu\rho}\right)\nonumber\\
        & \phantom{=}
            \phantom{-}\,\,\left[\Phi_+(v\cdot x)-\frac{x^2}{2(v\cdot x)^2}\Phi_-^{\text{t3}}(v\cdot x)+\frac{x^2}{2(v\cdot x)}\Phi_-^{\text{WW}\prime}(v\cdot x)\right.\nonumber\\
        & \phantom{=}
            \left.\phantom{-}\,\,\phantom{\left(\right.}-\frac{x^2}{4(v\cdot x)}\Phi_+^{\prime}(v\cdot x)+\frac{x^2}{4}\bar{\Psi}_4^+(v\cdot x)\right]\,.
\end{align}
In the above, we introduce an LCDA in position space that renders the above expression very compact,
\begin{equation}
\label{eq:framework:higher-twist:Psi4}
    \bar{\Psi}_4^+(z)
        \equiv \int_0^1 du \ \Psi_4^+(z, u z)\,,
    \quad\text{with}\quad
    \Psi_4^+(z_1, z_2)
        \equiv \big[\Psi_4+\tilde{\Psi}_4\big](z_1,z_2)+\big[\Psi_4+\tilde{\Psi}_4\big](z_2,z_1)\,.
\end{equation}
This form also facilitates expressing the higher-twist contributions at order $1/(v\cdot q)$ as dispersive integrals. 
In order to transform to the momentum-space representation, we use integral identities which are collected in \cref{app:details:identities}.
Performing the Fourier transform, projecting onto our basis, defining $\hat{q}^2 \equiv q^2 / (2 v\cdot q)$ and expanding in $1/(v\cdot q)$,
we obtain expressions for each of the form factors at order $\text{LP}+1/(v\cdot q)$.
These expression involve only three of the LCDAs: $\phi_+$, $\phi_-^{\text{t3}}$ and $\bar{\psi}_4^+$.

We organise our result for the $1/(v\cdot q)$ corrections in terms of two-particle twist-2 and twist-3 and three-particle twist-4 contributions as
\begin{equation}
    F_i\big|_{1/(v\cdot q)}
        =   F_i\big|_{1/(v\cdot q)}^\text{2pt,tw2}
            + F_i\big|_{1/(v\cdot q)}^\text{2pt,tw3}
            + F_i\big|_{1/(v\cdot q)}^\text{3pt,tw4}\,.
\end{equation} 
On carrying out the calculation, we find these three contributions to the form factors to be:
\begin{align}
\label{eq:framework:vq+2pt+tw2}  
    F_1\big|_{1/(v\cdot q)}^\text{2pt,tw2}
        & = -\frac{1}{2}F_2\big|_{1/(v\cdot q)}^\text{2pt,tw2} = F_4\big|_{1/(v\cdot q)}^\text{2pt,tw2}
        \\
    \nonumber
        & = \frac{Q_u f_B M_B}{4(v\cdot q)^2} \left[3\int_0^\infty \frac{d\omega}{\omega-\hat{q}^2}\,\omega\,\phi_+(\omega)+\int_0^\infty \frac{d\omega}{\omega-\hat{q}^2}\omega^2\phi^\prime_+(\omega)\right.\\
    \nonumber    &    \phantom{= \frac{Q_u f_B M_B}{4(v\cdot q)^2}}\quad\left.-2\int_0^\infty \frac{d\omega}{\omega-\hat{q}^2}\,\omega\int_\omega^\infty d\rho\frac{\phi_+(\rho)}{\rho}\right]\,,
        \\
\label{eq:framework:vq+2pt+tw3}
    F_1\big|_{1/(v\cdot q)}^\text{2pt,tw3}
        & = -\frac{1}{2}F_2\big|_{1/(v\cdot q)}^\text{2pt,tw3}=F_4\big|_{1/(v\cdot q)}^\text{2pt,tw3}
        \\ 
    \nonumber
        & = \frac{Q_u f_B M_B}{4(v\cdot q)^2}\left[2\int_0^\infty \frac{d\omega}{\omega-\hat{q}^2} \bar{\phi}_-^{\text{t3}}(\omega)\right]\,,
        \\
\label{eq:framework:vq+3pt+tw4}
    F_1\big|_{1/(v\cdot q)}^\text{3pt,tw4}
        & = -\frac{1}{2}F_2\big|_{1/(v\cdot q)}^\text{3pt,tw4}=F_4\big|_{1/(v\cdot q)}^\text{3pt,tw4}
        \\
    \nonumber
        & = \frac{Q_u f_B M_B}{4(v\cdot q)^2} \left[(-1)\int_0^\infty \frac{d\omega}{\omega-\hat{q}^2}\bar{\psi}_4^{+\prime}(\omega)\right]\,.
\end{align}
Here the prime indicates a derivative with respect to $\omega$, and in the twist-3 contribution we use the abbreviation
\begin{equation}
    \label{eq:framework:phimt3}
    \bar{\phi}_-^{(j)}(\omega)\equiv\int_\omega^\infty d\rho \ \phi_-^{(j)}(\rho)
\end{equation}
where the label $j$ (if present) is either ``WW'' or ``t3''.
We note that all $1/(v\cdot q)$ power correction are ``universal'', \ie~they contribute equally up to constant factors.
Our results generalize those presented in eqs.~(3.5) and (3.6) of ref.~\cite{Beneke:2018wjp} to arbitrary $q^2$.
We reproduce the previous results in the limit $q^2 \to 0$.\\

As indicated above, at this order in the power expansion the calculation of genuine $1/m_b$ corrections is also necessary.
On one hand, these corrections occur when the photon is emitted from the heavy quark. In this case, we obtain
\begin{equation}
    \label{eq:framework:twist-2:Qb}
    \begin{aligned}
        F_1\big|_{1/m_b}^{Q_b}(v\cdot q) & = \frac{Q_b f_B M_B}{2m_b(v\cdot q)}\,,                          &
        F_2\big|_{1/m_b}^{Q_b}(v\cdot q) & = -\frac{4}{3}\frac{Q_b f_B \bar{\Lambda}}{m_b(v\cdot q)}\,,     \\
        F_3\big|_{1/m_b}^{Q_b}(v\cdot q) & = 0 + \order{\alpha_s \frac{\Lambda}{m_b(v\cdot q)^2}},          &
        F_4\big|_{1/m_b}^{Q_b}(v\cdot q) & = \frac{Q_b f_B M_B}{2m_b(v\cdot q)}\,.
    \end{aligned}
\end{equation}
On the other hand,
there are genuine $1/m_b$ power corrections emerging from a covariant derivative acting on the $h_v$ field.
To calculate these correction to the form factor, we use the $1/m_b$ expression for the hadronic tensor
\begin{align}
    \label{eq:framework:higher-twist-mb:TH}
    T_{\text{H},1/m_b}^{\mu\nu}
        & = \frac{Q_u f_B M_B}{4\pi^2 m_b}\int\ d^4x\ \frac{e^{iq\cdot x}}{x^4}v_\rho x_\sigma\left(g^{\mu\nu}g^{\rho\sigma}+i\epsilon^{\mu\nu\rho\sigma}-g^{\mu\rho}g^{\nu\sigma}-g^{\mu\sigma}g^{\nu\rho}\right)\nonumber\\
        & \phantom{=}
        \,\,\,\left\lbrace \left(\omega-\bar{\Lambda}\right)\Phi_+(v\cdot x)-\bar{\Phi}_-(v\cdot x)-2i\left(v\cdot x\right)\bar{\Phi}_3(v\cdot x)\right\rbrace
\end{align}
where we use $\bar{\Lambda} = M_B - m_b$.
We further introduce two LCDAs in position space that render the above expression very compact. These are
\begin{equation}
        \bar{\Phi}_-(z)
            \equiv
                \int_{0}^{\infty} d\omega\, e^{-i\omega z} \bar{\phi}_-(\omega)\,,
        \quad\text{and}\quad
        \bar{\Phi}_3(z)
            \equiv
                \int_0^1 du \ (1-u)\Phi_3(z,uz)\,.
\end{equation}
Carrying out the Fourier transformation of \cref{eq:framework:higher-twist-mb:TH}, we obtain the $1/m_b$ power corrections as indicated in \cref{eq:framework:higher-twist:Fcontributions}. 

We organise our result for the $1/m_b$ corrections in terms of two particle twist-2 and three particle twist-3 contributions as follows:
\begin{equation}
    F_i\big|_{1/m_b}
        = F_i\big|_{1/m_b}^\text{2pt,tw2} + F_i\big|_{1/m_b}^\text{2pt,tw3} + F_i\big|_{1/m_b}^\text{3pt,tw3}\,.
\end{equation}
On carrying out the calculation, we find these three contributions to the form factors to be:
\begin{align}
\label{eq:framework:mb+2pt+tw2}
    F_1\big|_{1/m_b}^\text{2pt,tw2}
        & = -\frac{1}{2}F_2\big|_{1/m_b}^\text{2pt,tw2} = F_4\big|_{1/m_b}^\text{2pt,tw2}
        \\
    \nonumber
        & = \frac{Q_u f_B M_B}{4m_b(v\cdot q)} \left[\int_0^\infty  \frac{d\omega}{\omega-\hat{q}^2}(\bar{\Lambda}-\omega)\phi_+(\omega)\right.
        \\
    \nonumber
        & \phantom{= \frac{Q_u f_B M_B}{4m_b(v\cdot q)}\left[\right.} \left.\,\,-\int_0^\infty \frac{d\omega}{\omega-\hat{q}^2}\omega\int_\omega^\infty d\rho \ \frac{\phi_+(\rho)}{\rho} \right]\,,\\
\label{eq:framework:mb+2pt+tw3}
    F_1\big|_{1/m_b}^\text{2pt,tw3}
        & = -\frac{1}{2}F_2\big|_{1/m_b}^\text{2pt,tw3}=F_4\big|_{1/m_b}^\text{2pt,tw3}
        \\ 
    \nonumber
        & = \frac{Q_u f_B M_B}{4m_b(v\cdot q)}\left[\int_0^\infty \frac{d\omega}{\omega-\hat{q}^2} \bar{\phi}_-^{\text{t3}}(\omega)\right]\,,\\
\label{eq:framework:mb+3pt+tw3}
F_1\big|_{1/m_b}^\text{3pt,tw3}
        & = -\frac{1}{2}F_2\big|_{1/m_b}^\text{3pt,tw3}=F_4\big|_{1/m_b}^\text{3pt,tw3}
        \\
    \nonumber
        & = \frac{Q_u f_B M_B}{4m_b(v\cdot q)} \left[2\int_0^\infty \frac{d\omega}{\omega-\hat{q}^2}\bar{\phi}_3^\prime(\omega)\right]\,.
\end{align}
Our results generalize those presented in eqs.~(3.9) and (3.10) of ref.~\cite{Beneke:2018wjp} to arbitrary $q^2$.
We reproduce the previous results in the limit $q^2 \to 0$.

We note that our results in \cref{eq:framework:vq+2pt+tw2,eq:framework:mb+2pt+tw2}, while in dispersive form,
are only convergent integral representations when working to leading order in $\alpha_s$.
This is due to the onset of the radiative tail of the the leading-twist LCDA $\phi_+(\omega)$.
When using our results in an analysis that includes terms of order $\alpha_s/(v\cdot q)$ or $\alpha_s/m_b$,
one either needs to carefully track these divergent terms and ensure cancellation or use a subtracted dispersion relation.

\subsection{Soft contributions}
\label{sec:framework:soft-cont}

\begin{figure}
    \centering
    \begin{tikzpicture}
        \begin{feynman}
            \vertex (a);
            \vertex [below =1.7cm of a] (b);
            \vertex [left=of a] (i1) {$b$};
            \vertex [right=of a] (f1) {\(\gamma^*\)};
            \vertex [left=of b] (i2) {$\bar{u}$};
            \vertex [right=of b] (f2) {${W^-}^*$};
            \diagram*{
                        (i1) -- [fermion, very thick] (a) -- [photon, momentum={$q$}] (f1),
                        (a) -- [fermion, very thick, edge label=$b$] (b),
                        (i2) -- [anti fermion] (b) -- [photon] (f2)
                        };
            \draw [decoration={brace, mirror}, decorate] (i1.north west) -- (i2.south west) node [pos=0.5, left=5pt] {\large $B^-$};
            \draw[thick, dash pattern=on 5pt off 3pt] (-0.5,0.5) arc[start angle=180, end angle=270, radius=1cm];
        \end{feynman}
    \end{tikzpicture}
    \quad
    \begin{tikzpicture}
        \begin{feynman}
            \vertex (a);
            \vertex [below =1.7cm of a] (b);
            \vertex [left=of a] (i1) {$b$};
            \vertex [right=of a] (f1) {${W^-}^*$};
            \vertex [left=of b] (i2) {$\bar{u}$};
            \vertex [right=of b] (f2) {\(\gamma^*\)};
            \diagram*{
                        (i1) -- [fermion, very thick,] (a) -- [photon] (f1),
                        (b) -- [anti fermion, edge label=$\bar{u}$] (a),
                        (i2) -- [anti fermion] (b) -- [photon, momentum={$q$}] (f2)
                        };
            \draw [decoration={brace, mirror}, decorate] (i1.north west) -- (i2.south west) node [pos=0.5, left=5pt] {\large $B^-$};
            \draw[thick, dash pattern=on 5pt off 3pt] (0.5,-1.0) arc[start angle=90, end angle=180, radius=1cm];
        \end{feynman}
    \end{tikzpicture}
    \caption{%
        Sketches to illustrate the origin of soft contributions to the photoleptonic decay.
        In the left sketch, the photon is emitted from the $b$ quark line, which is a power correction
        to leading QCDF expression; here, soft contributions are not considered.
        In the right sketch, the photon is emitted from the $u$ quark line. This sketch represents
        a mechanism that gives rise to soft contributions, resonantly enhanced at $q^2 \simeq m_\rho^2$.
    }
    \label{fig:soft-cont:sketch}
\end{figure}
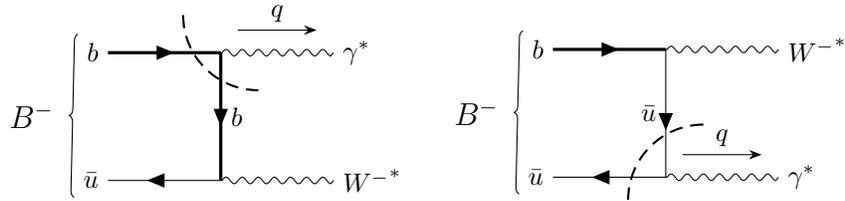

Despite the light-cone dominance, the time-ordered product giving rise to the hadronic matrix element in \cref{eq:intro:def-TH}
also receives soft contributions due to field configurations with $x^2 \simeq \Lambda_\text{had}^2$.
These soft contribution are not accessible within the framework of QCDF. However, an estimate using a light-cone sum rule setup is possible, which was first discussed for the case of an on-shell photon in ref.~\cite{Braun:2012kp}; further refinements followed in
refs.~\cite{Wang:2016qii,Beneke:2018wjp}.
Here, we revisit the derivation of the original light-cone sum rule setup for the general off-shell case. In the process, we critically
assess the impact of working with a basis of form factors free of kinematic singularities, which is a pre-requisite for the application
of dispersive methods such as light-cone sum rules.

The soft-contributions arising from interactions at $x^2 \simeq \Lambda_\text{had}^2$ cannot be captured in the framework
of QCDF. They therefore inhibit our ability to use $B^- \to \lbrace \gamma, \ell^{\prime-}\ell^{\prime+}\rbrace \ell\bar\nu$
data to extract information~\cite{Mandal:2023lhp,Lughausen:2024pmk} on the light-cone distribution amplitudes, either from
experimental measurements~\cite{LHCb:2018jvy}
or from lattice QCD simulations~\cite{Giusti:2023pot,Giusti:2025ibe}.
The biggest concern here is the creation of an intermediate on-shell $u\bar{u}$ pair at timelike $q^2$,
which is resonantly enhanced as $q^2$ approaches $m_\rho^2$ or the mass of any other unflavoured hadronic resonance with
the same quantum numbers as the photon; see \cref{fig:soft-cont:sketch} for a sketch of how this contribution arises.
This resonant enhancement is not restricted to purely timelike $q^2$. Instead, one can expect a substantial ``tail"
of the soft contribution also affecting $q^2 < m_\rho^2$.
On the other hand, for the purpose of dealing with soft contributions, the emission of the photon from the $b$ quark line is not a cause of concern.
This follows since a $b\bar{b}$ pair can only go on-shell at $q^2 \simeq 4m_b^2$, far beyond the kinematic boundaries of the
physical decay process.
Therefore, we only investigate soft contributions due to emission from the $u$ quark line.
We follow the original derivation for estimating these soft contributions as discussed in ref.~\cite{Braun:2012kp}
and start by considering their dispersion relation at fixed $k^2$,
\begin{equation}
    \label{dispersion}
    F_i(k^2,q^2)
        = \frac{f_\rho \mathcal{F}_i^{B\to\rho}(k^2)}{m_\rho^2 - q^2}
        + \frac{f_\omega \mathcal{F}_i^{B\to\omega}(k^2)}{m_\omega^2 - q^2}
        + \frac{1}{\pi}\int_{{s_0}}^\infty d s \, \frac{\Im\left\lbrace F_i(k^2, s)\right\rbrace}{s-q^2}\,.
\end{equation}
Here, the first two terms isolate the production of an intermediate on-shell $\rho$ or $\omega$ meson in a vector-meson dominance model,
whereas the last term corresponds to the production of all other intermediate on-shell multi-meson states above an effective threshold $s_0$.
The resonance terms are of particular interest, since they can be assumed to encode the brunt of the soft contributions.
The calligraphic $\mathcal{F}_i$ symbolize a linear combination of the respective $B\to \rho$ and $B\to \omega$ form factors.
Note that this representation of the form factors requires the absence of spurious kinematic singularities.
The presence of such spurious singularities would lead to additional terms in the dispersion integral.
Hence, this representation should ideally be used only in singularity-free basis of form factors proposed in ref.~\cite{Kurten:2022zuy}.
The broad nature of the $\rho$ leads to the expectation of a non-negligible tail, \ie~substantial contributions for $k^2$ values
significantly smaller than $M_\rho^2$, while the effects of the narrow $\omega$ resonance
will be more localized. We therefore concentrate on the $\rho$ contribution and drop the $\omega$-resonance term.

The QCDF results for the form factors can also be expressed in the form of a dispersive integral,
\begin{equation}\label{HQEdispersion}
    F_i(k^2,q^2)\big|_\text{QCDF}
        = \frac{1}{\pi}\int_0^\infty d s \, \frac{\Im\left\lbrace F_i^\text{QCDF}(k^2, s)\right\rbrace}{s-q^2}\,.
\end{equation}
The dispersive representation of the results at leading power \cref{eq:framework:twist-2:def-hc-kernels} and
our new results beyond leading power as provided in
\cref{eq:framework:vq+2pt+tw2,eq:framework:vq+2pt+tw3,eq:framework:vq+3pt+tw4,eq:framework:mb+2pt+tw2,eq:framework:mb+3pt+tw3}
makes it possible to read off
their imaginary parts by identifying $s$ with $(n_+q) \omega$.
For large values of $s$ above the effective threshold, one can now assume quark-hadron duality to hold locally~\cite{Braun:2012kp},
\begin{equation}\label{duality}
    \Im\left\lbrace F_i(k^2, s)\right\rbrace\simeq\Im\left\lbrace F_i^\text{QCDF}(k^2, s)\right\rbrace\quad\text{for }s>s_0\, ,
\end{equation}
\ie~assume that the partonic and hadronic contributions coincide~\cite{Colangelo:2000dp}.
Using this assumption, one equates \cref{dispersion} and \cref{HQEdispersion} to extract the resonance terms (\ie~the expected dominant soft contributions)
\begin{equation}
    \frac{f_\rho \mathcal{F}_i^{B\to\rho}(k^2)}{m_\rho^2 - q^2}
    = \frac{1}{\pi}\int_0^{s_0} d s \, \frac{\Im\left\lbrace F_i^\text{QCDF}(k^2, s)\right\rbrace}{s-q^2}\, .
\end{equation}
Note here that the characterisation of the resonance terms as being  dominant does not follow from a strict power-counting argument.

To reduce the sensitivity to the approximation in \cref{duality} and simultaneously suppress higher-twist contributions to the light-cone OPE results,
ref.~\cite{Braun:2012kp} suggests applying a Borel transformation to isolate the $B\to\rho$ form factor:
\begin{equation}
    \label{eq:soft:rhoFF}
    f_\rho \mathcal{F}_i^{B\to\rho}(k^2)=\frac{1}{\pi}\int_0^{s_0} d s \, \Im\left\lbrace F_i^\text{QCDF}(k^2, s)\right\rbrace e^{-\left(s-m_\rho^2\right)/M^2}\, .
\end{equation}
Inserting \cref{eq:soft:rhoFF} into \cref{dispersion} one obtains
\begin{equation}
    \begin{aligned}
        F_i(k^2,q^2)
        = & \frac{1}{\pi}\int_0^\infty d s \, \frac{\Im\left\lbrace F_i^\text{QCDF}(k^2, s)\right\rbrace}{s-q^2}\\
        & + \frac{1}{\pi}\int_0^{s_0} d s \, \frac{\Im\left\lbrace F_i^\text{QCDF}(k^2, s)\right\rbrace}{s-q^2}\left[\frac{s-q^2}{m_\rho^2-q^2}e^{-\left(s-m_\rho^2\right)/M^2}-1\right]\, ,
    \end{aligned}
\end{equation}
where, following ref.~\cite{Braun:2012kp}, we identify the first term as the leading contribution and the second term as the soft correction.
In terms of $\omega$ and $v\cdot q$ we obtain:
\begin{equation}
    \label{eq:leadplussoft}
    \begin{aligned}
        F_i(k^2,q^2)
        = & \frac{1}{\pi}\int_0^\infty d \omega \, \frac{\Im\left\lbrace F_i^\text{QCDF}(k^2, 2(v\cdot q)\omega)\right\rbrace}{\omega-\hat{q}^2}\\
        & + \frac{1}{\pi}\int_0^{\omega_0} d \omega \, \frac{\Im\left\lbrace F_i^\text{QCDF}(k^2, 2(v\cdot q)\omega)\right\rbrace}{\omega-\hat{q}^2}
        \left[\frac{\omega-\hat{q}^2}{\hat{m}_\rho^2-\hat{q}^2}e^{-2(v\cdot q)\left(\omega-\hat{m}_\rho^2\right)/M^2}-1\right]\,,
    \end{aligned}
\end{equation}
where $\omega_0\equiv s_0/2(v\cdot q)$ and $\hat{m}_\rho^2\equiv m_\rho^2/2(v\cdot q)$.
Reading off the imaginary parts of the form factors from \cref{eq:framework:vq+2pt+tw2,eq:framework:vq+2pt+tw3,eq:framework:vq+3pt+tw4,eq:framework:mb+2pt+tw2,eq:framework:mb+2pt+tw3,eq:framework:mb+3pt+tw3},
we obtain
\begin{equation}
    \begin{aligned}
        \Im\left\lbrace F_1\big|_{1/(v\cdot q)}^\text{2pt,tw2}\right\rbrace = & -\frac{1}{2}\Im\left\lbrace F_2\big|_{1/(v\cdot q)}^\text{2pt,tw2}\right\rbrace = \Im\left\lbrace F_4\big|_{1/(v\cdot q)}^\text{2pt,tw2}\right\rbrace\\
            = & \pi\frac{Q_u f_B M_B}{4(v\cdot q)^2} \left(3\omega \phi_+(\omega)+\omega^2\phi'_+(\omega)-2\omega\int_\omega^\infty d\rho \ \frac{\phi_+(\rho)}{\rho}\right)\\
        \Im\left\lbrace F_1\big|_{1/(v\cdot q)}^\text{2pt,tw3}\right\rbrace = & -\frac{1}{2}\Im\left\lbrace F_2\big|_{1/(v\cdot q)}^\text{2pt,tw3}\right\rbrace = \Im\left\lbrace F_4\big|_{1/(v\cdot q)}^\text{2pt,tw3}\right\rbrace\\
            = & \pi\frac{Q_u f_B M_B}{4(v\cdot q)^2} 2\int_\omega^\infty d\rho \ \phi_-^{\text{t3}}(\rho)\\
        \Im\left\lbrace F_1\big|_{1/(v\cdot q)}^\text{3pt,tw4}\right\rbrace = & -\frac{1}{2}\Im\left\lbrace F_2\big|_{1/(v\cdot q)}^\text{3pt,tw4}\right\rbrace = \Im\left\lbrace F_4\big|_{1/(v\cdot q)}^\text{3pt,tw4}\right\rbrace\\
            = & \pi\frac{Q_u f_B M_B}{4(v\cdot q)^2} \left(-1\right)\bar{\psi}_4^{+\prime}(\omega)\\
        \Im\left\lbrace F_1\big|_{1/m_b}^\text{2pt,tw2}\right\rbrace = & -\frac{1}{2}\Im\left\lbrace F_2\big|_{1/m_b}^\text{2pt,tw2}\right\rbrace = \Im\left\lbrace F_4\big|_{1/m_b}^\text{2pt,tw2}\right\rbrace\\ 
            = & \pi\frac{Q_u f_B M_B}{4m_b(v\cdot q)} \left[\left(\bar{\Lambda}-\omega\right)\phi_+(\omega)-\omega\int_\omega^\infty d\rho \ \frac{\phi_+(\rho)}{\rho}\right]\\
        \Im\left\lbrace F_1\big|_{1/m_b}^\text{2pt,tw3}\right\rbrace = & -\frac{1}{2}\Im\left\lbrace F_2\big|_{1/m_b}^\text{2pt,tw3}\right\rbrace = \Im\left\lbrace F_4\big|_{1/m_b}^\text{2pt,tw3}\right\rbrace\\
            = & \pi\frac{Q_u f_B M_B}{4m_b(v\cdot q)}\int_\omega^\infty d\rho \ \phi_-^{\text{t3}}(\rho)\\
        \Im\left\lbrace F_1\big|_{1/m_b}^\text{3pt,tw3}\right\rbrace = & -\frac{1}{2}\Im\left\lbrace F_2\big|_{1/m_b}^\text{3pt,tw3}\right\rbrace = \Im\left\lbrace F_4\big|_{1/m_b}^\text{3pt,tw3}\right\rbrace\\
            = & \pi\frac{Q_u f_B M_B}{4m_b(v\cdot q)} 2\bar{\phi}_3^\prime(\omega)\,.
    \end{aligned}
\end{equation}
This represents the central result of this work, allowing soft contributions due to higher-twist LCDAs  to be accounted for
in a consistent framework when matching the theoretical expressions to external numerical inputs at $q^2 \neq 0$.

\section{Numerical results}
\label{sec:numerics}

\begin{table}[t]
    \centering
    \begin{tabular}{c c c|c c c}
        \toprule
        Parameter  & Value         & Ref.                             & Parameter   & Value          & Ref.\\
        \midrule
        $M_B$      & $5.28\,\GeV$  & \cite{ParticleDataGroup:2024cfk} & $m_b$       & $4.8\,\GeV$    & \cite{ParticleDataGroup:2024cfk} \\
        $m_\rho$   & $775\,\MeV$   & \cite{ParticleDataGroup:2024cfk} & $\mu$       & $1.5\,\GeV$    & \cite{Beneke:2011nf} \\
        $f_B$      & $192\,\MeV$   & \cite{ParticleDataGroup:2024cfk} & $\lambda_B$ & $350\,\MeV$    & \cite{Beneke:2018wjp} \\
        $s_0$      & $1.5\,\GeV^2$ & \cite{Beneke:2018wjp}            & $M^2$       & $1.25\,\GeV^2$ & \cite{Beneke:2018wjp} \\
        $\mu_{h1}$ & $m_b$         & \cite{Beneke:2011nf}             & $\mu_{h2}$  & $m_b$          & \cite{Beneke:2011nf} \\
        \bottomrule
    \end{tabular}
    \caption{%
        Central values of the parameters used in the calculation of the soft corrections and the leading factorizable contribution.
    }
    \label{tab:numerics:parameters}
\end{table}
In this section, we quantify the size of the soft contributions obtained from the dispersive representation described in \cref{eq:leadplussoft}.
To do so, we need to make a choice for the kinematic variables: the photon virtuality $q^2$, and the momentum transfer to the
lepton-neutrino pair $k^2$.
To ensure that the expansion in $1/(v\cdot q)$ that underlies the QCDF framework converges,
we need to maintain a minimum photon energy of $v\cdot q \simeq 1.5\, \GeV$.
We further briefly investigate the dependence of the magnitude of the soft contributions relative to the QCDF contributions
as a function of $q^2$.
To achieve this, we investigate two values of $q^2$: the on-shell point $q^2 = 0$ and a point corresponding to a spacelike
photon with $q^2 = -2\,\GeV^2$. In combination with the minimum photon energy of $\simeq 1.5\,\GeV$, this constrains
$k^2$ to our nominal interval $0 \leq k^2 \leq 10\,\GeV^2$.
All numerical evaluations in this section make use of the input parameters collected in \cref{tab:numerics:parameters}.
For a representative numerical evaluation, we need to choose a model for the LCDAs.
For the leading-twist LCDA, we use the exponential model, which arises as a limiting case of a more systematic parametrisation~\cite{Feldmann:2022uok}.
This model reads
\begin{equation}
    \phi_+(\omega) = \frac{\omega}{\omega_0^2}\, e^{-\omega/\omega_0} \equiv \omega f(\omega)\,,
\end{equation}
where we use $\omega_0 = \lambda_B= 350~\MeV$ and introduce the profile function $f(\omega)$~\cite{Beneke:2018wjp}.
Using the profile function and applying tree-level relations between the light-cone operators, the authors of ref.~\cite{Beneke:2018wjp}
arrive at
\begin{equation}
    \phi_-^\text{t3}(\omega)
        =\frac{1}{6} \varkappa\left(\lambda_E^2-\lambda_H^2\right)\left[\omega^2 f^{\prime}(\omega)+4 \omega f(\omega)-2 \int_\omega^{\infty} d \rho f(\rho)\right]\,,
\end{equation}
with
\begin{equation}
    \varkappa^{-1}
        \equiv \frac{1}{6} \int_0^{\infty} d\omega \ \omega^3 f(\omega)=\bar{\Lambda}^2+\frac{1}{6}\left(2 \lambda_E^2+\lambda_H^2\right)
        \,.
\end{equation}
For the numerical interpretation of our results, we further require models for the LCDAs $\bar{\phi}_3(\omega)$ and $\bar{\psi}_4^+(\omega)$.
Following the approach of ref.~\cite{Beneke:2018wjp}, we obtain for these LCDAs the profile function expressions
\begin{equation}
    \begin{aligned}
        \bar{\phi}_3(\omega)
            & = -\frac{1}{2}\varkappa\left(\lambda_E^2-\lambda_H^2\right)\left[\int_0^\omega d\omega_1\int_{\omega-\omega_1}^\omega d\omega_2 \ \left(\omega_1+\omega_2-\omega\right)\left(\omega_1+\frac{\omega_2^2}{\omega_1}\right)f^\prime(\omega_1+\omega_2)\right.\\
            & \phantom{= -\frac{1}{2}\varkappa\left(\lambda_E^2-\lambda_H^2\right)\left[\right.}\left.+\int_0^\omega d\omega_1\int_\omega^\infty d\omega_2 \ \left(\omega_1+\omega_2-\omega\right)\omega_1f^\prime(\omega_1+\omega_2)\right.\\
            & \phantom{= -\frac{1}{2}\varkappa\left(\lambda_E^2-\lambda_H^2\right)\left[\right.}\left.+\int_\omega^\infty d\omega_1\int_0^\omega d\omega_2 \ \left(\omega_1+\omega_2-\omega\right)\frac{\omega_2^2}{\omega_1}f^\prime(\omega_1+\omega_2)\right]\,,
            \\
        \bar{\psi}_4^+(\omega)
            & = 2\varkappa\left(\lambda_E^2+\lambda_H^2\right)\int_0^\omega d\omega_1 \ \omega_1\int_{\omega-\omega_1}^\infty d\omega_2 \ f(\omega_1+\omega_2)\,.
    \end{aligned}
\end{equation}

\begin{figure}[t]
    \centering
    \includegraphics[width=0.45\textwidth]{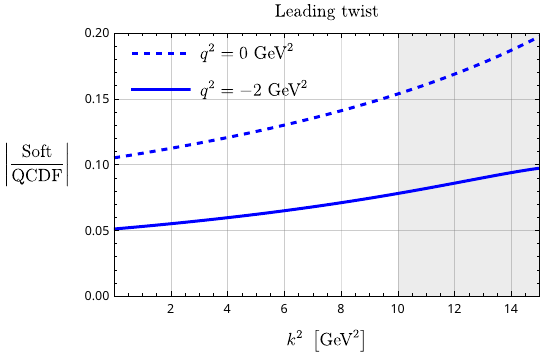}
    \qquad
    \includegraphics[width=0.45\textwidth]{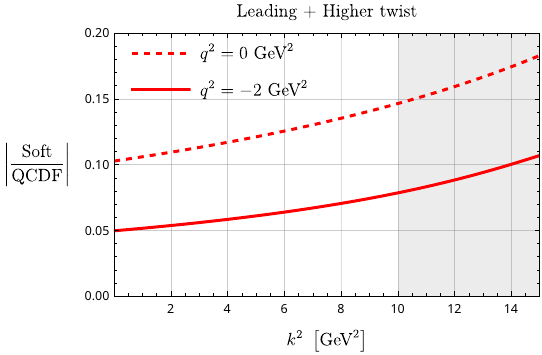}
    \caption{%
        The soft contributions for
        the form factor $F_4$,
        normalized to the QCDF result.
        We show the ratio for two fixed values of $q^2$ and as a function of $k^2$. The figure on the left includes only the leading power contribution while the figure on the right include all next-to-leading power contributions.}
    \label{fig:numerics:ratio-soft-v-collinear}
\end{figure}

We conduct our numerical study using $F_4$ serves as a representative result. The (almost) universal nature of the power correction ensures that
the our results for the form factors $F_1$ and $F_2$ are identical, up to constant factors. We further remind the reader that the form factor $F_3$ only arises as a power correction and does not contribute to experimentally accessible observables; it will therefore not
be discussed here any further.\\
We begin with plotting the form factor $F_4$ as a function of $k^2$ at the two specified values of $q^2$, presenting
these two plots based in \cref{fig:numerics:ratio-soft-v-collinear}. One the left-hand side,
we plot the ratio of soft contributions over QCDF contributions when limiting ourselves to the leading-twist
LCDA $\phi_+$, \ie~using results available in the literature prior to this work.
Based on this plot, we find that for $q^2 = 0$ the soft contributions can contribute as much as $16\%$ of the QCDF contributions.
This constitutes a sizeable contribution, drawing into doubt whether a determination of the LCDA parameters
at this kinematic point is possible with controllable systematic uncertainties.
Moving into the spacelike region, \ie~to $q^2 = -2\,\GeV^2$, the soft contributions in the nominal $k^2$ interval are significantly reduced.
We find that they reach their maximal value of $\simeq 8\%$ at $k^2 = 10\,\GeV^2$ and are significantly smaller in the remainder of the
interval.
Hence, when inferring the $B$-meson LCDA parameters, we find that matching the theory expressions for the form factor
to external inputs (such as experimental or lattice QCD determinations) at a mildly spacelike point reduces the systematic uncertainties.
\\
On the right-hand plot of \cref{fig:numerics:ratio-soft-v-collinear}, we repeat this exercise including also the QCDF and soft contributions
arising from the higher-twist LCDAs $\phi_-^\text{t3}$, $\bar{\phi}_3$, and $\bar{\psi}_4^+$.
For $q^2 = 0$, the picture does not change qualitatively. The soft contributions are substantial and increase with increasing $k^2$ up 
to $\simeq 15\%$.
For $q^2 = -2\,\GeV^2$, there is no qualitative change either.
The maximal relative contribution due to the soft contributions still occurs at the right end of the nominal $k^2$ interval
and reaches about $\simeq 8\%$.

\renewcommand{\arraystretch}{1.3}
\begin{table}[t]
    \centering
    \begin{tabular}{c c c c c c}
        \toprule
        $q^2$ [$\text{ GeV}^2$]
            & $F_4$
            & $\phi_+$
            & $\phi_-^\text{t3}$
            & $\bar{\phi}_3$
            & $\bar{\psi}_4^+$
            \\
        \midrule
        \multicolumn{6}{c}{QCDF contributions}\\
        \midrule
        \phantom{-}0
            & 0.436
            & 105.6\%
            & -0.6\%
            & -0.5\%
            & -4.5\%
            \\
        -2
            & 0.177
            & 102.8\%
            & -0.3\%
            & -0.3\%
            & -2.2\%
            \\
        \midrule
        \multicolumn{6}{c}{soft contributions}\\
        \midrule
        \phantom{-}0
            & -0.048
            & 108.7\%
            & -1.0\%
            & -0.5\%
            & -7.2\%
            \\
        -2
            & -0.007
            & 106.5\%
            & -0.7\%
            & -0.7\%
            & -5.1\%
            \\
        \bottomrule
    \end{tabular}
    \caption{%
        Representative breakdown of the various QCDF and soft contributions for the form factor $F_4$ at $k^2=2\,\GeV^2$.
        Moving from the on-shell photon to a mildly spacelike photon reduced the impact of the soft contributions by a factor $7$.
    }
    \label{tab:numerics:relative-contributions}
\end{table}

To assess the relative importance of the individual contribution to the form factor $F_4$, we disentangle them in \cref{tab:numerics:relative-contributions}.
There, we separate the form factor into its individual QCDF and soft contributions, organised by the respective LCDAs as introduced
in \cref{sec:framework:higher-twist}.
For the QCDF contributions, our findings are summarized as follows. At both $q^2$ points, the leading-twist LCDA $\phi_+$ dominates
the numerical results for the form factor. The higher-twist LCDAs lower the form factor by between $3\%$ and $5\%$, depending on the
value of $q^2$. Moving to mildly spacelike $q^2$ achieves a relative reduction of all higher-twist terms by a factor of $\simeq 2$.
At the same time, the over all result for the form factor is reduced by a factor of $\simeq 2.6$.
\\
On the other hand, we find the soft contributions to reduce more strongly than the QCDF contributions when we make $q^2$ more spacelike. In our example, this
reduction corresponds to a factor of $\simeq 18$. At the same time, we find that the relative impact of the soft contribution due to
the higher-twist LCDAs is increased.
\\

Overall, our numerical study indicates that working at sufficiently spacelike $q^2$ provides a pragmatic handle to reduce the impact of the
soft contributions. This means that, using a spacelike subtraction point in the dispersion relations, one is able to
determine the LCDA parameters from experimental or theoretical inputs for the form factors at timelike $q^2$ with controllable systematic uncertainties.

\section{Conclusion}
\label{sec:conclusions}

The decays $B^-\to \gamma \ell^-\bar\nu$ and $B^-\to \ell'^+\ell'^- \ell^-\bar\nu$ provide the opportunity to
probe the substructure of the $B$ meson by confronting measurements of the decay rates with theoretical predictions.
In this article, we have investigated the four physical form factors for $B^-\to \gamma^*$ transitions within the framework of QCD factorization (QCDF),
taking care to carry out this investigation for a basis of form factors that is free of spurious kinematic singularities.
We provide, for the first time, results at non-zero $q^2$ that include power corrections at orders $1/m_b$ and $1/(v\cdot q)$, relative to the leading terms.
Our results agree with those for the form factor to an on-shell photon already available in the literature.
We have taken care to express our results in the form of a dispersion relation.

Besides the dominant hard-collinear contributions arising in QCDF, the form factors receive so-called soft contributions
that cannot be computed at present.
Following previous work, we provide an estimate of these contributions within a light-cone sum rule setup.
This setup makes use of a dispersion relation for the form factors, and the expressions within this framework reuse the dispersive integrals
and integration kernels from our QCDF results.
Importantly, our work generalises the existing estimates based on $B$-meson light-cone distribution amplitudes available in the literature,
which only account for two of the four form factors.

We provide benchmark numerical results for two photon virtualities, illustrating that the soft contributions represent a substantial part of the full result
for an on-shell photon. However, we find that shifting the photon virtuality mildly into the spacelike region, the soft contribution
reduces much more strongly than the total form factor result, to $\leq 8\%$ of the total form factor.
We therefore conclude that determinations of the $B$-meson LCDA parameters
from external inputs, such as experimental or lattice QCD results, should ideally be carried out at spacelike photon virtualities;
a virtuality as small as $q ^2 \simeq -2\,\GeV^2$ suffices for this purpose.

\section*{Acknowledgments}

We are very grateful to Vladimir Braun and Yao Ji for helpful discussions and comparisons of partial results for the
on-shell form factors.
DvD~acknowledges support by the UK Science and Technology Facilities Council (grant numbers ST/V003941/1 and ST/X003167/1).

\appendix

\section{Details}
\label{app:details}

\subsection{Final state radiation}
\label{app:details:FSR}

The final state radiation (FSR) contribution is commonly included through the matrix element $\tilde{T}_\mathrm{FSR}^\mu$, defined via
\begin{equation}
    Q_{\ell} \tilde{T}_{\mathrm{FSR}}^\mu\left(p_{\ell}, p_\nu, q\right)
        = -i f_B p_\nu \int d^4 x \ e^{i q\cdot x}\matrixel{\ell^{-} \bar{\nu}_{\ell}}{\mathrm{T}\left\{J_{\mathrm{em}}^\mu(x) J_{\mathrm{L}}^\nu(0)\right\}}{0}
        \,,
\end{equation}
where $J_\mathrm{em}^\mu$ is the electromagnetic current as defined in \cref{eq:intro:Jem}, and $J_L^\nu$ is the leptonic current
introduced in \cref{eq:framework:def-amp}.
It is convenient to define instead an FSR tensor~\cite{Kurten:2022zuy}:
\begin{equation}
    \TFSR^{\mu\nu} \left(p_{\ell}, p_\nu, q\right)
        =   f_B\left[
                g^{\mu \nu} + \frac{2 p_{\ell}^\mu p_{\ell}^\nu + p_{\ell}^\mu q^\nu + q^\mu p_{\ell}^\nu-\left(p_{\ell} \cdot q\right) g^{\mu \nu}
                + \mathrm{i} \epsilon^{\mu \nu \rho \sigma}\left(p_{\ell}\right)_\rho q_\sigma}{\left(p_{\ell}+q\right)^2-m_{\ell}^2}
            \right]
\end{equation}
which takes the role of $\tilde{T}_\mathrm{FSR}^\mu$ only when contracted with the leptonic matrix element $L^\nu \equiv \bra{\ell^-\bar\nu}J_{L}^{\nu}\ket{0}$,
\begin{equation}
    e \varepsilon^*_\mu\left[
        Q_{B^-} T_\mathrm{H}^{\mu\nu} L_\nu + Q_\ell \tilde{T}_\mathrm{FSR}^\mu
    \right]
        =
            e Q_{B^-} \varepsilon^*_\mu\left[
                T_\mathrm{H}^{\mu\nu} + T_\mathrm{FSR}^{\mu\nu}
            \right] L_\nu 
\end{equation}
Only the full amplitude including both the hadronic tensor $T_\mathrm{H}$ and the FSR tensor $T_\mathrm{FSR}$
is gauge invariant and therefore fulfils the Ward identity
\begin{equation}
    q_\mu\left[
        T_{\mathrm{H}}^{\mu \nu}(k, q) + T_{\mathrm{FSR}}^{\mu \nu}\left(p_{\ell}, p_\nu, q\right)
    \right] L_\nu
        = 0
        \,.
\end{equation}
Individually, the two tensors instead fulfil the inhomogeneous Ward identities
\begin{equation}
    \begin{aligned}
        q_\mu T_{\mathrm{H}}^{\mu \nu}(k, q)
            & = -f_B(k+q)^\nu\,,
            \\
        q_\mu T_{\mathrm{FSR}}^{\mu \nu}\left(p_{\ell}, p_\nu, q\right)
            & = +f_B(k+q)^\nu\,.
    \end{aligned}    
\end{equation}

\subsection{Singularity-free hadronic tensor}
\label{app:details:singularity-free-hadronic-tensor}

Following ref.~\cite{Kurten:2022zuy}, we split the hadronic tensor into a homogeneous part and an inhomogeneous part by means of
$T_{\mathrm{H}}^{\mu \nu}(k, q)=T_{\mathrm{H}, \text {hom.}}^{\mu \nu}(k, q)+T_{\mathrm{H}, \text {inhom.}}^{\mu \nu}(k, q)$,
which fulfil
\begin{equation}
    \begin{aligned}
        q_\mu T_{\mathrm{H}, \text {hom.}}^{\mu \nu}(k, q) & =0, \\
        q_\mu T_{\mathrm{H}, \text {inhom.}}^{\mu \nu}(k, q) & =-f_B(k+q)^\nu .
    \end{aligned}
\end{equation}
To ensure that the homogeneous part does not suffer from spurious kinematic singularities, we use~\cite{Kurten:2022zuy}
\begin{equation}
    T_{\mathrm{H}, \text { inhom. }}^{\mu \nu}(k, q)
        = -f_B\left[g^{\mu \nu}+\frac{\left(2 k^\mu+q^\mu\right) k^\nu}{2(k \cdot q)+q^2}\right]
        \,.
\end{equation}

\subsection{Projectors onto the form factors}
\label{app:details:projectors}

Here, we collect expressions for the projector $P_i^{\mu \nu}(k, q)$ that fulfil $P_{i \mu \nu}(k, q) T_{\mathrm{H}}^{\mu \nu}(k, q) = F_i\left(k^2, q^2\right)$ for $i=1, \ldots, 4$ as provided in eq.~(C1) of ref.~\cite{Kurten:2022zuy}:
\begin{equation}
    \begin{aligned}
        \frac{1}{M_B} P_1^{\mu \nu}(k, q)
            & = \frac{k \cdot q}{2\left[(k \cdot q)^2-k^2 q^2\right]} g^{\mu \nu} + \frac{3 q^2(k \cdot q)}{2\left[(k \cdot q)^2-k^2 q^2\right]^2} k^\mu k^\nu
                -\frac{(k \cdot q)^2+2 k^2 q^2}{2\left[(k \cdot q)^2-k^2 q^2\right]^2} k^\mu q^\nu
            \\
            & \phantom{=}\,\,\,-\frac{3(k \cdot q)^2}{2\left[(k \cdot q)^2-k^2 q^2\right]^2} q^\mu k^\nu+\frac{3 k^2(k \cdot q)}{2\left[(k \cdot q)^2-k^2 q^2\right]^2} q^\mu q^\nu\,,
            \\
        \frac{1}{M_B} P_2^{\mu \nu}(k, q)
            & = \frac{k^2}{2\left[(k \cdot q)^2-k^2 q^2\right]} g^{\mu \nu} + \frac{2(k \cdot q)^2+k^2 q^2}{2\left[(k \cdot q)^2-k^2 q^2\right]^2} k^\mu k^\nu
                -\frac{3 k^2(k \cdot q)}{2\left[(k \cdot q)^2-k^2 q^2\right]^2} k^\mu q^\nu
            \\
            & \phantom{=}\,\,\, -\frac{3 k^2(k \cdot q)}{2\left[(k \cdot q)^2-k^2 q^2\right]^2} q^\mu k^\nu + \frac{3 k^4}{2\left[(k \cdot q)^2-k^2 q^2\right]^2} q^\mu q^\nu\,,
            \\
        \frac{1}{M_B} P_3^{\mu \nu}(k, q)
            & = \frac{1}{(k \cdot q)^2-k^2 q^2} k^\mu k^\nu - \frac{k^2}{\left[(k \cdot q)^2-k^2 q^2\right]\left[2(k \cdot q)+q^2\right]} q^\mu k^\nu
            \\
            & \phantom{=}\,\,\, -\frac{k^2}{\left[(k \cdot q)^2-k^2 q^2\right]\left[2(k \cdot q)+q^2\right]} q^\mu q^\nu\,,
            \\
        \frac{1}{M_B} P_4^{\mu \nu}(k, q)
            & = -\frac{\mathrm{i}}{2\left[(k \cdot q)^2-k^2 q^2\right]} \epsilon^{\mu \nu \rho \sigma} k_\rho q_\sigma\,.
    \end{aligned}
\end{equation}

\subsection{LCDA definitions}
\label{app:details:lcda-definitions}

Throughout this work, we use the common definitions for the two-particle and three-particle LCDAs~\cite{Braun:2017liq}.
The two-particle LCDAs are defined as
\begin{equation}
    \begin{aligned}
        \matrixel{0}{\bar{q}(x) \Gamma[x, 0] h_v(0)}{\bar{B}(v)} =& -\frac{i}{2} F_B \operatorname{Tr}\left\lbrace\gamma_5 \Gamma P_{+}\right\rbrace\left(\Phi_{+}(\omega)+x^2 G_{+}(\omega)\right) \\
        &  +\frac{i}{4} F_B \operatorname{Tr}\left\lbrace\gamma_5 \Gamma P_{+} \slashed{x}\right\rbrace \frac{1}{v x}\left(\left[\Phi_{+}-\Phi_{-}\right](\omega)+x^2\left[G_{+}-G_{-}\right](\omega)\right)\,.    
\end{aligned}
\end{equation}
The three-particle LCDAs are defined as
\begin{equation}
    \begin{aligned}
        \hspace{-.2cm}\matrixel{0}{\bar{q}\left(n z_1\right) g G_{\mu \nu}\left(n z_2\right) \Gamma h_v(0)}{\bar{B}(v)} =&\frac{1}{2} F_B(\mu) \operatorname{Tr}\Big\lbrace\gamma _ { 5 } \Gamma P _ { + } \Big[\left(v_\mu \gamma_\nu-v_\nu \gamma_\mu\right)\left[\Psi_A-\Psi_V\right]-i \sigma_{\mu \nu} \Psi_V \\
        &-\left(n_\mu v_\nu-n_\nu v_\mu\right) X_A+\left(n_\mu \gamma_\nu-n_\nu \gamma_\mu\right)\left[W+Y_A\right] \\
        &-i \epsilon_{\mu \nu \alpha \beta} n^\alpha v^\beta \gamma_5 \widetilde{X}_A+i \epsilon_{\mu \nu \alpha \beta} n^\alpha \gamma^\beta \gamma_5 \widetilde{Y}_A\\
        &-\left(n_\mu v_\nu-n_\nu v_\mu\right) \slashed{n} W+\left(n_\mu \gamma_\nu-n_\nu \gamma_\mu\right) \slashed{n} Z\Big]\Big\rbrace\left(z_1, z_2 ; \mu\right)\,.
\end{aligned}
\end{equation}
The LCDAs defined above do not have a definite twist. Ref.~\cite{Braun:2017liq} provides the necessary information to
translate them to LCDAs of definite twist as follows
\begin{equation}
    \begin{aligned}
        \Psi_A & =\frac{1}{2}[\Phi_3+\Phi_4]\left(z_1, z_2;\mu\right), \\
        \Psi_V & =\frac{1}{2}[-\Phi_3+\Phi_4]\left(z_1, z_2;\mu\right), \\
        X_A & =\frac{1}{2}[-\Phi_3-\Phi_4+2 \Psi_4]\left(z_1, z_2;\mu\right), \\
        Y_A & =\frac{1}{2}[-\Phi_3-\Phi_4+\Psi_4-\Psi_5]\left(z_1, z_2;\mu\right), \\
        \tilde{X}_A & =\frac{1}{2}[-\Phi_3+\Phi_4-2 \tilde{\Psi}_4]\left(z_1, z_2;\mu\right), \\
        \tilde{Y}_A & =\frac{1}{2}[-\Phi_3+\Phi_4-\tilde{\Psi}_4+\tilde{\Psi}_5]\left(z_1, z_2;\mu\right), \\
        W & =\frac{1}{2}[\Phi_4-\Psi_4-\tilde{\Psi}_4+\tilde{\Phi}_5+\Psi_5+\tilde{\Psi}_5]\left(z_1, z_2;\mu\right), \\
        Z & =\frac{1}{4}[-\Phi_3+\Phi_4-2 \tilde{\Psi}_4+\tilde{\Phi}_5+2 \tilde{\Psi}_5-\Phi_6]\left(z_1, z_2;\mu\right)\,.
\end{aligned}
\end{equation}
In this work, we only include LCDAs up to and including twist 4.

\subsection{LCDA identities}
\label{app:details:identities}

In deriving the next-to-leading power contributions to the form factors, we make use of two relations between the various LCDAs.
These relations ensure that we can express our results both in terms of a profile function and in a dispersive form.
We therefore rewrite $\big[\Psi_4-\tilde{\Psi}_4\big]^{\text{t4}}(z,uz)$ by applying the tree-level identity
\begin{equation}\label{eq:Psi4}
    \int_0^1 du \ \big[\Psi_4-\tilde{\Psi}_4\big]^{\text{t4}}(z,uz)
        = -\frac{1}{4z}\left[\Phi_-^{\text{WW}\prime}(z)+\Phi'_+(z)+2i\bar{\Lambda}\Phi_+(z)\right]-\bar{\Psi}_4^+(z)\, ,
\end{equation}
where the prime indicates the derivative.
In the above, $\bar{\Psi}_4^+(z)$ is defined in \cref{eq:framework:higher-twist:Psi4}.
We further eliminate $G_+^\text{WW}(z)$ by applying the tree-level identity
\begin{equation}\label{eq:GWWpl}
    G_+^{\text{WW}}(z)
        = \frac{1}{4z}\left[\Phi_-^{\text{WW}\prime}(z)-2\Phi'_+(z)-2i\bar{\Lambda}\Phi_+(z)\right]\,.
\end{equation}
To transform to the momentum-space representation, we use the following integral identities
\begin{align}
\label{eq:intidLCDA}
    \frac{\Phi_-^{\text{t3}}(z)}{z^2}
        & = -\int_0^\infty d\omega\int_0^\omega d\eta\int_0^\eta d\rho \ e^{-i\omega z}\phi_-^{\text{t3}}(\omega)\,,
        \\
    \frac{\Phi_-^{\text{WW}\prime}(z)}{2z}-\frac{\Phi_+^{\prime}(z)}{4z}
        & = \frac{1}{4}\int_0^\infty d\omega \ e^{-i\omega z}\omega^2\phi_-^{\text{WW}}(\omega)\,.
\end{align}

\bibliographystyle{jhep} 
\bibliography{references.bib}

\end{document}